\documentclass[prl, twocolumn, showpacs, showkeys, preprintnumbers, amsmath, amssymb]{revtex4}
\usepackage{graphicx}                        % Include figure files
\usepackage{bm}                              % Bold math

\def\abs#1{\ifmmode \left \vert #1 \right \vert \else $\left \vert #1 \right \vert$ \fi}

\begin{document}

% ===============================================
%                                               %
%                     TITLE                     %
%                                               %
% ===============================================
\title{Reconstruction of Rb-Rb inter-atomic potential from ultracold Bose-gas collision}

\author{D{\'a}niel Schumayer}                                                                  % \email{dschumayer@physics.otago.ac.nz}
\email{dschumayer@physics.otago.ac.nz}
\affiliation{Department of Physics, University of Otago, New Zealand}

\author{Oliver Melchert}                                                                        % \email{melchert@theorie.physik.uni-goettingen.de}
\author{Werner Scheid}                                                                          % \email{Werner.Scheid@theo.physik.uni-giessen.de}
\affiliation{Institute for Theoretical Physics, Justus-Liebig Universit{\"a}t, Germany.}

\author{Barnab{\'a}s Apagyi}                                                                    %\email{apagyi@phy.bme.hu}
\affiliation{Department of Physics, Budapest University of Technology and Economics, Hungary}

\date{\today}

\begin{abstract}
    Scattering phase shifts obtained from $^{87}$Rb Bose-gas collision
    experiments are used to reconstruct effective potentials
    resulting, self-consistently, in the same scattering events
    observed in the experiments at a particular energy. We have found
    that the interaction strength close to the origin suddenly
    changes from repulsion to attraction when the collision energy
    crosses, from below, the $l=2$ shape resonance position at
    $E\approx 275$ $\mu$K. This observation may be utilized in outlining
    future Bose-gas collision experiments.
\end{abstract}

% PACS, the Physics and Astronomy Classification Scheme
% --------------------------------------------------------------------
% 34.20.-b    Inter-atomic and intermolecular potentials and forces,
%             potential energy surfaces for collisions
% 34.20.Cf    Inter-atomic potentials and forces
% 34.50.-s    Scattering of atoms and molecules
% 82.20.Fd    Collision theories; trajectory models
% 02.30.Zz    Inverse problems
% 03.75.-b    Matter waves (for atom interferometry techniques, 
%             see 39.20.+q - in atomic and molecular physics)
\maketitle

% ===============================================
%                                               %
%                 INTRODUCTION                  %
%                                               %
% ===============================================
\paragraph{Introduction \label{sec:Introduction}}
Scattering and collision processes have always been one of the most
important tools providing information on interaction between quantum
systems or particles at all lengths and energy scales. In the quantum
regime, we usually cannot perform direct experiments with atoms,
electrons or other particles. What we can do is merely to reconstruct
scattering events and thereby the interaction properties of the
constituents from experimentally observed quantities of the collision.
At high energy, particle-colliders serve to probe Nature at the
smallest distances. At the lower end of the energy scale, we can cool
down atoms to the ultracold regime ($<$ 1$\mu$K) \cite{StevenChu1986,
Dalibard1989} and form a Bose-Einstein condensate (BEC)
\cite{Anderson1995} to explore the low-energy properties of atomic
interaction.

At low temperatures, collisions play a pivotal role, mainly in
affecting the static and dynamic properties of the condensate:
stability, lifetime and thermalization rate. Furthermore, it is much
simpler to describe the dominant binary collisions theoretically than
at higher energies, because inelastic processes are usually
negligible. In early BEC experiments, the $s$ partial wave and the
associated scattering length $a_{\mathrm{s}}$ \cite{Wigner1948} have
been sufficient to characterize the low-energy properties of the
condensate. However, to produce molecular condensate from alkali
atoms, or to tune the interaction strength via Fesh{\-}bach
resonances, or to discover effects beyond the mean-field theory, we
need to know the interatomic potential between atoms more precisely.
Both theoretical calculations \cite{Julienne1998, Kempen2002,
Raoult2004, Durr2005} and experimental efforts \cite{Wynar2000,
Roberts2001, Seto2000} have been devoted to this aim, e.g., the
authors of Ref.~\cite{Seto2000} presented high-precision fluorescence
data and accurately fitted parameters of the electronic ground state
of Rb$_{2}$. Theoretical calculations usually involve some
experimental quantities, like scattering lengths \cite{Kempen2002},
positions of Feshbach resonance~\cite{Kempen2002, Marte2002} or Raman
transition rates~\cite{Julienne1998}. These results are then compared
with other data, and if the correspondence is not satisfactory, the
potential used in the calculation is iteratively adjusted.

Our approach is conceptually different. We want to characterize or
even reconstruct the inter-atomic potential from scattering phase
shifts, $\eta_{l}$, by using the inverse scattering method. Inverse
scattering theories (see \cite{Chadan1989, Apagyi1997} and references
therein) provide two types of potentials. One type is derived from
phase shifts measured at a particular angular momentum but for all
energies. The other type of potentials is calculated from all partial
wave phase shifts corresponding to a fixed energy. The underlying idea
of both types of inverse scattering theory is the assumption that
there exists an ``effective'' spherical potential which is the cause
of the observed scattering events. In this Letter, we employ the
fixed-energy inverse scattering theory and use experimental phase
shifts $\eta_{l}^{\mathrm{exp}}$ derived from Bose-gas collision
experiments \cite{Thomas2004, Buggle2004}. Because the experimental
phase shifts are largely riddled with errors, we rely upon the phase
shifts $\eta_{l}^{\mathrm{J}} \approx \eta_{l}^{\mathrm{exp}}$ of
Julienne's coupled channel calculation \footnote{private
communication}, by which the BEC collision experiment
\cite{Thomas2004} has been interpreted. Notice that these phase shifts
stem from a coupled-channel calculation and reproduce the well-known
$d$--resonance in the $l=2$ partial wave at collision energy $E\approx
275$ $\mu$K. Although this resonance is considered as a shape
resonance of the triplet $^{87}$Rb $-$ $^{87}$Rb potential
\footnote{We thank Eite Tiesinga for illuminating discussion.},
without its inclusion one cannot explain \cite{Durr2005} important
coupled channels effects like tunability of scattering length, which
is a typical Fesh{\-}bach resonance effect. We demonstrate that the
collisional phase shifts derived from experiments also contain such
information, if we consider them at fixed energy but all partial waves.
Note that this consideration is precisely the situation which occurs
in collision experiments.

In the following, we briefly review the Bose-gas collision experiments
\cite{Thomas2004, Buggle2004} which provide the input data for the
inversion. Thereafter we collect all the the formulas necessary to
reconstruct potentials from phase shift data. In discussing the
results, first we present the $^{87}$Rb-$^{87}$Rb inversion potentials
extracted from measurements carried out in the low collision energy
domain 100-200 $\mu$K, then we exhibit these effective interactions in
the resonance region around 275 $\mu$K. Here a sudden change from 
repulsion to attraction of potentials can be observed and interpreted
as the coupled channel effect mentioned above. Finally, we show that
well beyond the resonance region, between 600-1200 $\mu$K where only
calculated phase shifts are available, the potentials appear to be
independent of energy and retain their attractive character.

% ===============================================
%                                               %
%                   THE MODEL                   %
%                                               %
% ===============================================
\paragraph{Review of BEC collision experiments}
In the experiment of Ref.~\cite{Thomas2004}, $^{87}$Rb atoms were
prepared in the $\left \vert F = 2, m_{F}=2 \right \rangle$ hyperfine
state and precooled to about 12 $\mu$K. Afterward the cloud of atoms
was adiabatically split into two pieces by an emerging potential
barrier in the middle of the trapping potential. Thereby the atom
cloud experienced a double well potential and was divided into two
parts separated by 4.3 mm. The barrier height was several times higher
than the chemical potential, therefore the clouds were isolated and
could be further cooled down to 225 nK. After such a preparation, the
double-well potential was ramped down to a single-well harmonic
potential characterized by the angular frequencies $\omega_{r} = 2\pi
\times 155$ Hz and $\omega_{z} = 2 \pi \times 12$ Hz, and was kept
constant during the measurement. As a result, the stacks of atoms were
accelerating towards each other and finally they collided with a
relative velocity $v$. The collision energy range
$E_{\mathrm{kin}}/k_{B} = \mu v^{2} /2k_{B}$ was between 87 and 553
$\mu$K calculated in the centre-of-mass frame. Due to symmetrical
collision of identical bosons, the $p$ partial wave was prohibited.
Following the impact, the atoms were scattered and were moving in the
trap until their maximum extension. At this moment, an absorption
image was taken with a resonant light shone onto the clouds,
perpendicular to the scattering axis. Although this image was a
two-dimensional projection of a three-dimensional density, it was
possible to reconstruct \cite{Thomas2004} the full tomographical
information taking into account the cylindrical symmetry.

In a similar experiment \cite{Buggle2004}, it has been demonstrated
that it is not necessary to keep the atoms in the trap all the time
after the collision. The analysis of the observed scattering halo
provides all the data. Calculating the density in small angular
sectors yields us the angular scattering distribution of the halo,
which is directly proportional to the differential cross section
$\sigma (\theta)$. Taking into account only the $l$=0, 2 partial waves
and fitting the analytical expression 
\begin{equation*}
   \sigma (\theta)
   = 
   \frac{1}{k^2}
   \abs{   \left ( e^{2i\eta_{0}^{\mathrm{exp}}} - 1 \right ) + 
         5 \left ( e^{2i\eta_{2}^{\mathrm{exp}}} - 1 \right ) 
           \frac{3\cos^{2}(\theta)-1}{2}
       }^{2}
\end{equation*}
to the measured angular distribution finally provides the experimental
phase shifts $\eta_{0}^{\mathrm{exp}}$ and $\eta_{2}^{\mathrm{exp}}$.
These latter phase shifts constitute the input for our inverse
calculations.

% ===============================================
%                                               %
%               INVERSE SCATTERING              %
%                                               %
% ===============================================
\paragraph{Inverse scattering formulas}
We employ the fixed energy inverse scattering method of Cox and
Thompson (CT) \cite{Cox1970} in order to derive model independent
potentials from the given phase shifts \cite{Thomas2004,Buggle2004}
corresponding to the measurements. The CT inversion method has a
number of useful properties \cite{Apagyi2003, Melchert2006}:
we may work with a finite set of $N$ experimental phase shifts and
obtain inversion potentials of non-zero first momentum, $\int{r V(r)
{\mathrm{d}}r \ne 0}$, and of finite value at the origin. Let us
denote the set of physical angular momenta $l$ by $S$ and the set of
unknown `shifted' angular momenta $L$ by $T$. The latter has to be
determined from the phase shifts. $T$ contains the same number $N$ of
elements as $S$, and the sets $S$ and $T$ are disjoint. The CT method
leads to a system of non-linear equations
\begin{equation} \label{Invscattering_eq1}
    e^{2i\delta_{l}} = \frac{1 + i{\mathcal{K}}^{+}_l}%
                            {1 - i{\mathcal{K}}^{-}_l}, 
\end{equation}
in which the input scattering phase shifts ($\delta_{l}$)
determine the `shifted' reactance matrix elements
defined as
\begin{equation} \label{Invscattering_eq2}
   {\mathcal{K}}_{l}^{\pm} 
   = 
   \sum_{L\in T ,l'\in S}{ {\mathcal{N}}_{lL} 
                           \left( {\mathcal{M}}^{-1} \right )_{Ll'}
                           e^{\pm i(l-l')\pi/2}
                         },
\end{equation}
with the square matrices
\begin{equation}
   \hspace*{-3mm}  
   \genfrac{ \{ }{ \} }{0pt}{}{\mathcal{N}}{\mathcal{M}}_{lL} 
   = 
   \frac{1}{L(L+1) - l(l+1)}
   \genfrac{ \{ }{ \} }{0pt}{0}{\sin{\left ((l-L)\pi/2 \right )}}%
                               {\cos{\left ( (l-L)\pi/2 \right )}}_{lL}
\end{equation}
containing the unknown $L-$values. Once the set $T$ is determined by
solving the highly nonlinear equation \eqref{Invscattering_eq1} for
the $L$s, we calculate coefficient functions $A_L(r)$ using the system
of linear equations
\begin{equation} \label{Invscattering_eq4}
   \sum_{L \in T}{ \frac{A_L(r)W[j_L(r),n_l(r)]}{l(l+1)-L(L+1)}}
   = 
   n_l(r),
\end{equation}
where $j_{L}$ and $n_{l}$ mean the spherical Bessel and Neumann
functions, respectively, and $W[a,b] = ab'-a'b$ denotes the
Wronski determinant. Next, we compute the transformation kernel
\begin{equation} \label{Invscattering_eq5}
   K(r,r')=\sum_{L\in T}A_{L}(r)j_L(r')
\end{equation}
from which the inversion potential is obtained as
\begin{equation} \label{Invscattering_eq6}
   V(r) = -\frac{2}{r} 
           \frac{\mathrm{d}}{{\mathrm{d}}r} 
           \frac{K(r,r)}{r}.
\end{equation}

% -----------------------------------------------------------------------------
% FIGURE: LOW ENERGY POTENTIALS
% -----------------------------------------------------------------------------
\begin{figure}[!hb]
   \includegraphics[width=60mm, angle=-90]{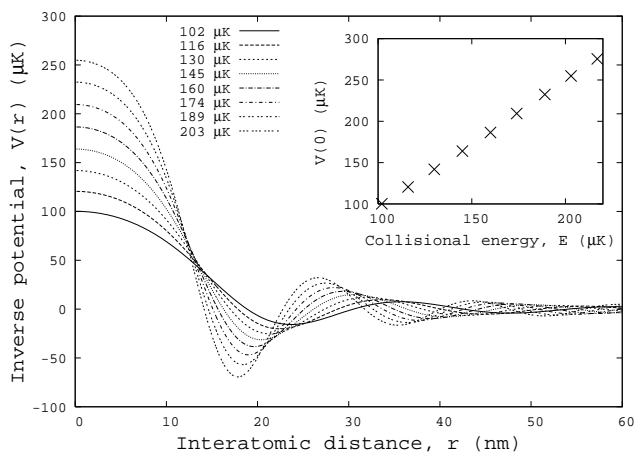}
   \caption{\label{fig:InversionPotentials_LowEnergy}
            Inverse potentials as functions of interatomic distance
            obtained from the $l=0$, $2$, $4$ scattering phase shifts
            within the range E=102-203 $\mu$K, below the $d$--resonance.
            Inset shows the central amplitude of the inverse potential
            as a function of collisional energy.}
\end{figure}

% ===============================================
%                                               %
%              RESULT AND DISCUSSION            %
%                                               %
% ===============================================
\paragraph{Results and discussion}
In Fig.~\ref{fig:InversionPotentials_LowEnergy} we present the results
for the energy range between 100-200 $\mu$K, which lies below the
characteristic $l=2$ resonance position of the $^{87}$Rb -- \!$^{87}$Rb
triplet scattering. The inverse potentials are repulsive at smaller
distances and oscillatory for larger relative coordinates. The
strength of the repulsion is approximately proportional to the
scattering energy, as is the attractive first minimum, the position of
which gets smaller values from 25 nm to 18 nm.

\begin{figure}[!hb]
   \includegraphics[trim=10 50 -3 22, clip=true, width=60mm, height=85mm, angle=-90]{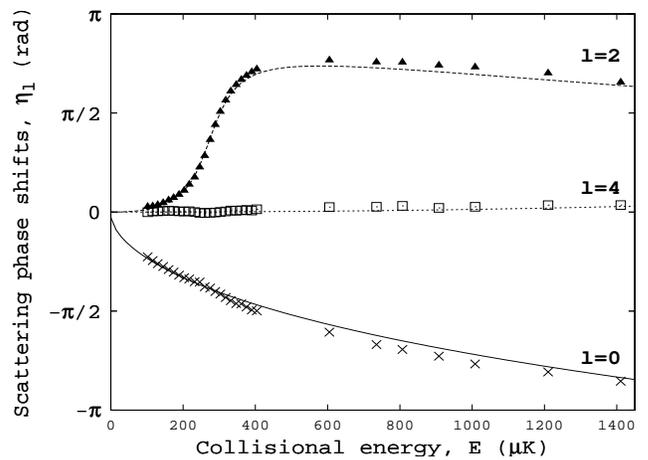}
   \caption{\label{fig:PhaseShiftsComparison}
            Phase shifts corresponding to the first three allowed partial
            waves ($l=0$, $2$, $4$) as functions of the collisional energy.
            Solid lines represent the original input data. Symbols $\times$,
            $\blacktriangle$ and $\boxdot$ stand for the phase shifts calculated
            from the inverse potentials shown in Figs.~\ref{fig:InversionPotentials_LowEnergy},
            \ref{fig:InversionPotentials_ResonanceEnergyRegion} and 
            \ref{fig:InversionPotential_FullEnergyRange}.}
\end{figure}

Not unexpectedly, the inversion potentials reproduce the input phase
shifts well within the considered energy region. This fact is
demonstrated in Fig.~\ref{fig:PhaseShiftsComparison}, where both input
and output phase shift values are exhibited. The phase shift
reproduction is rather accurate for the partial waves $l=0$, $2$, $4$
involved in the inversion procedure. But it must be so because the
only control over the potentials is the reproduction of the initial
data, since there is no free parameter in the inversion calculation
which is, in principle, unique \cite{Cox1970}. Numerical uncertainty
sometimes may lead to false potentials which can be recognized on
false reproduction of the input phase shifts. 

% -----------------------------------------------------------------------------
% FIGURE: PHASE COMPARISON
% -----------------------------------------------------------------------------
\begin{figure}[!hb]
   \includegraphics[width=60mm, angle=-90]{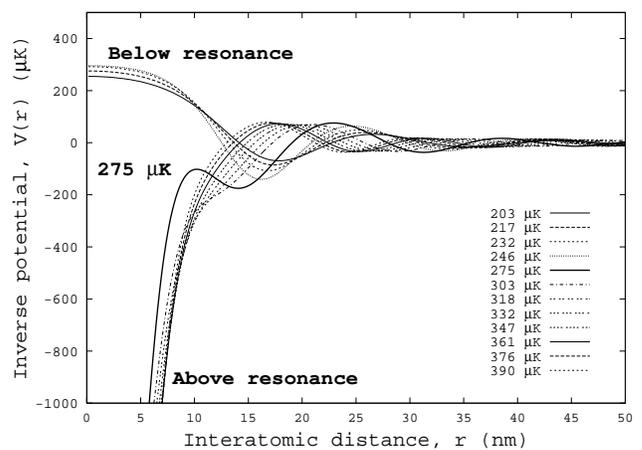}
   \caption{\label{fig:InversionPotentials_ResonanceEnergyRegion}
            Inverse potentials as functions of interatomic distance
            obtained from the $l=0$, $2$, $4$ scattering phase shifts
            around the $d$--resonance, $E$=200-400 $\mu$K.}
\end{figure}

It is therefore important that the excellent reproduction of input
phase shifts proceeds further on up to 400 $\mu$K, that is throughout
the whole $l=2$ resonance region. The corresponding inversion
potentials are exhibited in
Fig.~\ref{fig:InversionPotentials_ResonanceEnergyRegion}, and we
observe that these `on-resonance' potentials produce an abrupt change
of potential strength $V(0)$ from repulsion to attraction as the
collision energy crosses the $l=2$ resonance position at $ \approx
275$ $\mu$K. This behavior is quite unexpected in view of the input
phase shifts depicted by lines in
Fig.~\ref{fig:PhaseShiftsComparison}, which exhibits a smooth behavior
in the $s-$wave and a smooth shape resonance in the $d-$wave. Because
of this quite unusual behavior of the inversion potential we have
performed another control test of the results besides the comparison
of input and output phase shifts mentioned above. We have inverted
the original phase shifts of Ref.~\cite{Buggle2004} at two collision
energies $E=203$ and $447$ $\mu$K. The values of these measured phase
shifts differ slightly from the ones obtained by the coupled channel
calculation \cite{Thomas2004}. Therefore we can test at the same time
both the stability of the inversion procedure and the sensitivity of
the potential against small errors in phase shifts. The inversion
potentials obtained from the data of Ref.~\cite{Buggle2004} also
resulted in the sudden change and were practically the same as the
ones calculated from phase shifts of Ref.~\cite{Thomas2004}.

% -----------------------------------------------------------------------------
% FIGURE: POTENTIALS IN THE RESONANCE REGION
% -----------------------------------------------------------------------------
Let us look at the results from another viewpoint too. During the
collision process the energy is fixed by the experimental device and
the colliding atoms are approaching each other from asymptotic
distances to the smallest ranges. This is described in terms of
partial waves. How many partial waves are involved depends on the
details of the interaction of the colliding atoms, just as the very
numerical values of the partial wave phase shifts do depend on that.
In these Bose-gas experiments, it has been found that the first three
($l=0,2,4$) phase shifts have utmost importance at each fixed energy
and these phase shifts contain all the information about the
interaction of the colliding $^{87}$Rb gas particles. We have put
these three phase shifts belonging to each particular energy into the
inversion procedure, which provided potentials that produced the same
scattering events as those observed in the experiments. Since the
inversion procedure is unique \cite{Cox1970}, we may assume that the
inversion potentials are the effective potentials which govern the
collisions. 

\begin{figure}[!hb]
   \includegraphics[width=60mm, angle=-90]{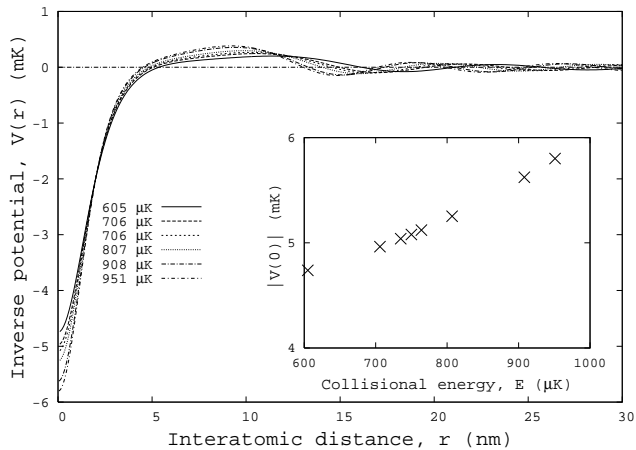}
   \caption{\label{fig:InversionPotential_FullEnergyRange}   
            Inverse potentials as functions of interatomic distance
            obtained from the $l=0$, $2$, $4$ scattering phase shifts
            within the range $E$=606-951 $\mu$K, above the
            $d$--resonance. Inset depicts the moduli of central
            amplitude of the inverse potentials as a function of
            collisional energy.
	       }
\end{figure}

In Fig.~\ref{fig:InversionPotential_FullEnergyRange} we show the
high-energy (well above the resonance) inversion potentials which do
not depend too much on the energy but retain their attractive
character.
% -----------------------------------------------------------------------------
% FIGURE: INVERSE POTENTIALS ABOVE THE RESONANCE
% -----------------------------------------------------------------------------

% ===============================================
%                                               %
%              RESULT AND DISCUSSION            %
%                                               %
% ===============================================
\paragraph{Summary}
The inversion potentials shown in
Figs.~\ref{fig:InversionPotentials_LowEnergy},
\ref{fig:InversionPotentials_ResonanceEnergyRegion} and
\ref{fig:InversionPotential_FullEnergyRange} can be viewed as
`effective' interactions which characterize the $^{87}$Rb interaction
in the considered energy range, although these Bose-gas collision
experiments can also be described by sophisticated coupled channels
calculation where the `potential' is represented by a potential
matrix. One or more elements of these potential matrix may become
dominant over the others at different energies. In case of the
$d$--resonance, the triplet potential plays a governing role in
explaining the observed cross section. In other cases, channel
couplings may destroy the dominant effects of a particular element of
the potential matrix. This might be the reason of the change in
nature of our potentials as we vary the collisional energy, since even
a weak interaction of different elements of the potential matrix can
cause resonances (e.g., shape- or Feshbach-resonance). Our effective
potentials thus enable intuitive interpretation of the interaction
between colliding $^{87}$Rb atoms and can be considered as a
one-channel mapping or a local and energy-dependent equivalent form of
the whole interaction matrix. 

Since the input data stem entirely from experimentally confirmed data
and the inverse procedure provides unique results, we do expect that
the sudden change of strength of the potentials can be observed and
utilized in future Bose-gas experiments. Moreover, the method of
inverse scattering may gain ground in the analysis of low-energy
collisions as it does in high energy physics.

% ===============================================
%                                               %
%              ACKNOWLEDGEMENT                  %
%                                               %
% ===============================================
\begin{acknowledgments}
   This work was supported by the Hungarian Scientific Research Fund,
   under contracts OTKA-T47035, T49571 and the MTA-DFG grant (436 UNG
   113/158). D. S. acknowledges the financial support from the Marsden
   Fund of the Royal Society of New Zealand.
\end{acknowledgments}

% ===============================================
%                                               %
%                 BIBLIOGRAPHY                  %
%                                               %
% ===============================================

% ********************************************************************
% *                          END OF ARTICLE                          *
% ********************************************************************
\end{document}